\documentclass[11pt]{article}

\usepackage{latexsym,amsmath}

\DeclareFontFamily{OT1}{rsfs10}{}
\DeclareFontShape{OT1}{rsfs10}{m}{n}{ <-> rsfs10 }{}
\DeclareMathAlphabet{\mathscript}{OT1}{rsfs10}{m}{n}

\numberwithin{equation}{section}

\newcommand{\ns}{\normalsize}
\newcommand{\pt}{\partial}

\newcommand{\be}{\begin{equation}}
\newcommand{\ee}{\end{equation}}

\newcommand{\bea}{\begin{eqnarray}}
\newcommand{\eea}{\end{eqnarray}}

\newcommand{\eval}[1]{\left.{#1}\right|}
\newcommand{\orbav}[1]{\big<{#1}\big>_{11}}
\newcommand{\tr}{\textrm{tr}}

\def\a{\alpha}

\def\d{\delta}
\def\e{\epsilon}

\def\f{\phi}

\def\k{\kappa}
\def\l{\lambda}

\def\o{\omega}
\def\p{\pi}

\def\r{\rho}

\begin{document}

%%%%%%%%%%%%%%%%%%%%%%%%%%%%%%%%%%%%%%%%%%%%%%%%%%%%%%%%%%%%%%%%%%%%%%

\begin{titlepage}

\vspace{-2cm}

\title{
   \hfill{\ns UPR-789T, PUPT-1757\\}
   \hfill{\ns hep-th/9801087\\[.5cm]}
   {\LARGE 
      The Ten-dimensional Effective Action of Strongly Coupled 
      Heterotic String Theory}}
\author{
   Andr\'e Lukas$^1$
      \setcounter{footnote}{0}\thanks{Supported in part by Deutsche
          Forschungsgemeinschaft (DFG).}~~,
   Burt A.~Ovrut$^1$
      \setcounter{footnote}{3}\thanks{Supported in part by a Senior 
          Alexander von Humboldt Award}~~and 
   Daniel Waldram$^2$\\[0.5cm]
   {\ns $^1$Department of Physics, University of Pennsylvania} \\
   {\ns Philadelphia, PA 19104--6396, USA}\\[0.3cm]
   {\ns $^2$Department of Physics}\\
   {\ns Joseph Henry Laboratories, Princeton University}\\
   {\ns Princeton, NJ 08544, USA}\\[0.3cm]}
\date{}

\maketitle

\begin{abstract}
We derive the ten-dimensional effective action of the strongly
coupled heterotic string as the low energy limit of M--theory on $S^1/Z_2$.
In contrast to a conventional dimensional reduction, it is necessary to
integrate out nontrivial heavy modes which arise from the sources
located on the orbifold fixed hyperplanes. This procedure,
characteristic of theories with dynamical boundaries, is illustrated
by a simple example. Using this method, we determine a complete set
of $R^4$, $F^2R^2$, and $F^4$ terms and the corresponding
Chern-Simons and Green-Schwarz terms in ten dimensions. As required by
anomaly cancelation and supersymmetry, these terms are found to
exactly coincide with their weakly coupled one-loop counterparts.
\end{abstract}

\thispagestyle{empty}

\end{titlepage}

%%%%%%%%%%%%%%%%%%%%%%%%%%%%%%%%%%%%%%%%%%%%%%%%%%%%%%%%%%%%%%%%%%%%%%%

\section{Introduction}

Almost two years ago, Ho\v{r}ava and Witten completed the cycle of
string theory dualities by relating the strongly coupled limit of the
$E_8\times E_8$ heterotic string to M--theory compactified on a
$S^1/Z_2$ orbifold~\cite{hw1,hw2}. The low-energy effective action of
M--theory is usually simply that of eleven-dimensional
supergravity. However, the presence of the orbifold projection means
that the gravitino fields are chiral
ten-dimensional fields on the two fixed hyperplanes of the
orbifold. In order to cancel the corresponding anomalies which appear
in fermion loops, it is necessary to introduce two sets of
ten-dimensional $E_8$ gauge fields, one on each fixed
hyperplane. Supersymmetry then requires that the bulk and hyperplane
theories are not independent. In particular, the gauge fields act as
magnetic sources for the four-form field strength of the bulk
supergravity, and as stress-energy sources for the graviton.  

When further compactified on a Calabi-Yau space, at tree level, the
strong limit provides a better match to the predicted four-dimensional
gravitational and grand-unified couplings than does the weakly coupled
heterotic string. Witten has shown that such compactifications exist~\cite{w},
although the internal space becomes distorted, while Ho\v{r}ava showed that
the theory has a topological version of gaugino
condensation~\cite{hor}. This has led to a number of papers
reconsidering basic four-dimensional string phenomenology in this new
limit~\cite{bd,aq1,kap,lln,ns,du1,choi,hp,aq2,lt,du2,efn,
low1,nan,ckm,low2,hp2}. 

To discuss the low-energy physics, one derives a four-dimensional
effective action by dimensional reduction. The full eleven-dimensional
effective action is an expansion in powers of the eleven-dimensional
gravitational coupling constant $\k$. The lowest dimension operators
of the four-dimensional theory then have a double expansion in terms of
the Calabi-Yau and orbifold sizes compared to the eleven-dimensional
Planck length $\k^{2/9}$. To zeroth order, the dimensional reduction to
four dimensions is simple because the source terms on the orbifold
fixed planes can be ignored. To this order, the derivation is identical
to reducing on a circle. However, to the next order the reduction is
more complicated. The presence of sources localized on the fixed
planes means that one cannot consistently take all fields to be
independent of the orbifold direction, as one would in a conventional
truncation on a circle. In order to match the boundary conditions
implied by the sources, the bulk fields must vary across the
interval. The inclusion of extra terms which arise in such a reduction
was an important ingredient in the derivation of the full
low-energy action, including some terms of order $\k^{4/3}$, given in
two previous papers~\cite{low1,low2}. 

The purpose of the present paper is to refine exactly how this
reduction works and to use it to calculate the ten-dimensional
effective action for the strongly coupled $E_8\times E_8$ heterotic
string. This is the limit where the orbifold interval remains large
compared to the eleven-dimensional Planck length but all fields are
assumed to have wavelengths much longer than the orbifold size. This
action can then be compared with the corresponding weakly coupled theory. 
In particular, we will concentrate on deriving the local higher-order
terms in the Riemann curvature $R_{AB}$ and gauge curvature $F_{AB}$,
up to quartic order. 

In general curvature terms enter the ten-dimensional action has in an
expansion of the form~\cite{ts} 
\begin{equation}
\label{effSexp}
   S = \int \sum_{n=1}^{\infty} a_n(\phi) A^n
\end{equation}
where $A$ represents either the Riemann or gauge curvature, and $\phi$
is the dilaton, which is the modulus of the compact eleventh dimension
in the strongly coupled theory. In the perturbative limit, in the
string frame, the coefficients can be expanded as
$a_n=b_{n,0}e^{-2\phi}+b_{n,1}+b_{n,2}e^{2\phi}+\cdots$. For the
lower-dimension terms, supersymmetry implies a number of strong
constraints on the form of $a_n$. Up to quadratic order, it is well
known that the only supersymmetric invariants are $e^{-2\phi}R$,
$e^{-2\phi}R^2$ and $e^{-2\phi}F^2$, where the last two terms must be
paired with Chern-Simons terms in the three-form field $H$~\cite{susyR2}. 
Furthermore there are no supersymmetric invariants of the form
$A^3$~\cite{R3}. At quartic order there are two types of
invariant~\cite{rsw,ts}. There is a parity-odd term
$t_8A^4-\frac{1}{2\sqrt{2}}\e^{(10)}BA^4$, with no dilaton
dependence and so with only a one-loop contribution in the weak
expansion. (For notation and a definition of $t_8$ see section 3
below.) There is also a parity-even term including $f(\phi)t_8t_8R^4$
and which does not include a coupling to $B$, which appears to allow
arbitrary dilaton dependence. It is terms of this form which appear as
stringy tree-level corrections in the weakly coupled limit. The
parity-odd invariants include the Green-Schwarz term required for
anomaly cancelation. This, in fact, then fixes the coefficients of
these terms with respect to the lower-dimension terms. Consequently,
ignoring the parity-even quartic terms, the form of the effective
action is completely fixed to this order by supersymmetry and anomaly
cancelation. (This is the simpler heterotic analog of the
non-renormalization of $R^4$ terms found in type II
theories~\cite{R4}.) In this paper we will show how these terms appear
when the strongly coupled theory is reduced to ten-dimensions. We will
not explicitly include the parity even terms, but will make some
comments about how they might arise. 

A calculation of some such terms and a discussion of
anomaly cancelation was first presented by Dudas and
Mourad~\cite{dm}, who noted that to obtain the full result it would be
necessary actually to integrate out the massive Kaluza-Klein modes. As we
will see, this is exactly the procedure we will perform. By clarifying
the form of the dimensional reduction, we will find that we can
reproduce the full parity-odd $R^4$, $R^2F^2$ and $F^4$ supersymmetric
invariants. While we will not explicitly consider other terms of the same
dimension but involving other fields in the supergravity and gauge
multiplets, we will find that the dimensional reduction procedure
provides a convenient way of deriving complete higher-order supersymmetric 
actions. The result that the $R^4$, $R^2F^2$ and $F^4$ terms do not
renormalize beyond one loop helps explain the result that the low-energy
four-dimensional effective actions derived in~\cite{low1,low2} have
the same form as those derived in the weakly coupled limit
including loop corrections on a large Calabi-Yau manifold. 

Let us end this introduction by summarizing our conventions. We denote
the coordinates in the eleven-dimensional spacetime $M_{11}$ by
$x^0,\ldots,x^9,x^{11}$ and the corresponding indices by
$I,J,K,\ldots=0,\ldots,9,11$. The orbifold $S^1/Z_2$ is chosen in the
$x^{11}$--direction, so we assume that $x^{11}\in [-\pi\r ,\pi\r ]$
with the endpoints identified as $x^{11}\sim x^{11}+2\pi\r$. The $Z_2$
symmetry acts as $x^{11}\rightarrow -x^{11}$. Then there exist two
ten-dimensional hyperplanes, $M_{10}^{(i)}$ with $i=1,2$, locally specified by
the conditions $x^{11}=0$ and $x^{11}=\pi\r$, which are fixed under
the action of the $Z_2$ symmetry. We will sometimes use the
``downstairs'' picture where the orbifold is considered as an interval
$x^{11}\in [0,\pi\r ]$ with the fixed hyperplanes forming boundaries
to the eleven-dimensional space. In the ``upstairs'' picture the
eleventh coordinate is considered as the full circle with singular
points at the fixed hyperplanes. We will use indices
$A,B,C,\ldots=0,\ldots,9$ to label the ten-dimensional coordinates. 

%%%%%%%%%%%%%%%%%%%%%%%%%%%%%%%%%%%%%%%%%%%%%%%%%%%%%%%%%%%%%%%%%%%%%%%%%%%

\section{Dimensional reduction with orbifold sources}

As we have stressed in the introduction, new features arise when
making a dimensional reduction on a $S^1/Z_2$ orbifold with
sources on the orbifold fixed planes. In this section we will show how
such a reduction can be performed consistently in the case when it is possible
to make a perturbative expansion in the strength of the
sources. Rather than consider the full eleven-dimensional description
of the strongly coupled heterotic string, we will illustrate the
issues involved in a simpler toy model with only a scalar field. 

Consider a scalar field $\phi$ in the bulk of the eleven-dimensional
spacetime, together with two sources $J^{(1)}$ and $J^{(2)}$ localized on the
two fixed hyperplanes of the orbifold. In M--theory on $S^1/Z_2$, the r\^ole
of $\f$ will be played by the eleven-dimensional metric and the three-form,
while the sources are provided by gauge fields and $R^2$ terms on the
orbifold fixed planes. Therefore, in general, the sources will be functionals
of fields living on the orbifold fixed planes.

Let us consider the following simple theory, in the upstairs picture,
with a standard kinetic term for $\phi$ and a linear coupling of
$\phi$ to the sources, 
\begin{equation}
\label{scalarS}
   S = - \int_{M_{11}} \frac{1}{2}\left(\partial\phi\right)^2
          - \int_{M_{10}^{(1)}} J^{(1)} \phi - \int_{M_{10}^{(2)}}
          J^{(2)} \phi \; ,
\end{equation}
which is sufficient to illustrate the main points. The scalar field must
have definite charge under the orbifold symmetry. Since the sources involve
the value of $\phi$ on the fixed planes, we must take $\phi$ to be even under
$Z_2$, so that $\phi(-x^{11})=\phi(x^{11})$.

The equation of motion for $\phi$ reads
\begin{equation}
\label{scalarEOM}
   \pt^2\phi = J^{(1)} \d(x^{11}) + J^{(2)} \d(x^{11}-\p\r)
\end{equation}
By Gauss's theorem, for a small volume intersecting the orbifold plane
we can integrate this equation to give, near $x^{11}=0$ and
$x^{11}=\p\r$ respectively
\begin{equation}
\label{nrbdry}
\begin{split}
   \pt_{11}\phi &= \frac{1}{2}J^{(1)}\e(x^{11}) + \ldots \\
       &= \frac{1}{2}J^{(2)}\e(x^{11}-\p\r) + \ldots
\end{split}
\end{equation}
where $\e(x)$ is the step function, equal to $1$ for $x>0$ and $-1$ for
$x<0$. To derive this expression, we have used that fact that
$\pt_{11}\phi$ must be odd under the $Z_2$ orbifold symmetry. The dots
represent terms which vanish at $x^{11}=0$ in the first line, or at
$x^{11}=\p\r$ in the second line. 

In the downstairs picture, the orbifold is an interval bounded by
the hyperplanes. Rather than having an equation of motion with delta
function sources, we then have the free equation $\pt^2\phi=0$
together with the boundary conditions, corresponding to the limiting
expressions for $\pt_{11}\phi$ as one approaches the boundaries, 
\begin{equation}
\label{scalarBC}
   \eval{n^{(1)}_I \pt^I \phi}_{M_{10}^{(1)}} = \frac{1}{2}J^{(1)} \qquad
   \eval{n^{(2)}_I \pt^I \phi}_{M_{10}^{(2)}} = -\frac{1}{2}J^{(2)} \; .
\end{equation}
Here $n^{(i)}$ are normal unit vectors to the two hyperplanes, pointing
in the direction of increasing $x^{11}$. They are introduced solely
in order to write these expressions in a covariant way. Note that the
second expression comes with a negative sign since it is evaluated
just to the left of the $x^{11}=\p\r$ orbifold plane, while the first
expression comes with a positive sign since it is evaluated just to
the right of the $x^{11}=0$ plane. 

In a conventional dimensional reduction, one makes a Fourier expansion
in the compact direction. The massive Kaluza-Klein modes have no
linear coupling to the massless modes and have masses set by the
size of the compact dimension. Thus at low energies they decouple and
the effective theory can be obtained by simply dropping them from the
action. More formally, this is equivalent to integrating them out at
tree-level. (Massive loops can however contribute, but since in this
paper we will be reducing what is already an effective action, we will
not be interested in this possibility.) It is clear from the $\phi$
equation of motion~\eqref{scalarEOM} that a similar truncation will
not work here. We might imagine trying to assume that both $\phi$ and
the sources $J^{(i)}$ are independent of the eleventh
coordinate. However, the delta functions in the sources mean that even
though $J^{(i)}$ are independent of $x^{11}$ we must have $\phi$
depending on the orbifold coordinate if we are to solve the equation
of motion. Equivalently, we clearly can not satisfy the boundary
conditions~\eqref{scalarBC} without $\phi$ depending on
$x^{11}$. There is no consistent solution where the massive modes are
set to zero. Instead we must consider more carefully what happens when
these modes are integrated out. 

First, we can consider expanding both $\phi$ and the total source in
the scalar equation of motion~\eqref{scalarEOM} in Fourier modes. We
write 
\begin{equation}
\label{scalarexp}
\begin{aligned}
   \phi &= \phi^{(0)} + \sum_n \tilde{\phi}^{(n)} \cos(nx^{11}/\r) 
       = \phi^{(0)} + \phi^{(B)} \\
   J &= J^{(1)} \d(x^{11}) + J^{(2)} \d(x^{11}-\p\r) 
       = J^{(0)} + \sum_n \tilde{J}^{(n)} \cos(nx^{11}/\r)
       = J^{(0)} + J^{(B)}\; .
\end{aligned}
\end{equation}
Since both $\phi$ and the total source $J$ must be even under the
$Z_2$ symmetry, only the cosine terms in the Fourier expansion
contribute. We can give an explicit form for the  Fourier components
$\tilde{J}^{(n)}$, but all we will require in what follows is that 
\begin{equation}
\label{Jzero}
\begin{aligned}
   J^{(0)} &= \frac{1}{2\p\r} \left( J^{(1)} + J^{(2)} \right) \\
   J^{(B)} &=  J^{(1)} \d(x^{11}) + J^{(2)} \d(x^{11}-\p\r) - 
       \frac{1}{2\p\r} \left(  J^{(1)} + J^{(2)} \right)\; .
\end{aligned}
\end{equation}
We note, confirming the discussion above, that even with $J^{(i)}$
independent of $x^{11}$, the massive mode source $J^{(B)}$ is
non-zero. By definition, with the eleven dimensional average given by
\begin{equation}
\label{orbav}
   \orbav{F} = \frac{1}{\pi\r}\int_{0}^{\pi\r}dx^{11}F
\end{equation}
the massive modes average to zero, $\orbav{\phi^{(B)}}=\orbav{J^{(B)}}=0$. 

We can now substitute these expansions in the action~\eqref{scalarS}
and integrate out the massive Kaluza-Klein modes. Rather than
performing this integration separately for each mode, it is easier to
integrate out $\phi^{(B)}$ as an eleven-dimensional field. Substituting
into the action we have
\begin{equation}
\label{scalarsubsS}
   S = - 2\p\r \int_{M_{10}} \left\{ \frac{1}{2}\left(\pt\phi^{(0)}\right)^2 
             + J^{(0)}\phi^{(0)} \right\}
       - \int_{M_{11}} \left\{ \frac{1}{2}\left(\pt\phi^{(B)}\right)^2
             + J^{(B)}\phi^{(B)} \right\}
\end{equation}
where $M_{11}=M_{10}\times{S^1/Z_2}$. Because the massless and
massive modes are orthogonal, they separate in the action. The massive
modes can now be integrated out. A simple Gaussian integration gives
\begin{equation}
\label{scalarintS}
   S = - 2\p\r \int_{M_{10}} \left\{ \frac{1}{2}\left(\pt\phi^{(0)}\right)^2 
             + J^{(0)}\phi^{(0)} \right\}
       - \int_{M_{11}} \frac{1}{2}J^{(B)}\Phi^{(B)}
\end{equation}
where $\Phi^{(B)}$ is the solution of the massive equation of motion
\begin{equation}
\label{massiveEOM}
   \pt^2\phi^{(B)} = J^{(B)} \quad \Leftrightarrow \quad
    \Phi^{(B)}(x)\equiv \phi^{(B)}(x) = \int_{x'} G(x-x')J^{(B)}(x')
\end{equation}
with $G(x-x')$ the eleven-dimensional Green's function. Using the
form of $J^{(0)}$ and $J^{(B)}$ given in~\eqref{Jzero} and the fact that
$\orbav{\Phi^{(B)}}$ is zero, we find that the action can be written
as the ten-dimensional action
\begin{multline}
\label{scalarten}
   S_{10} = - 2\p\r \int_{M_{10}} \left\{ 
           \frac{1}{2}\left(\pt\phi^{(0)}\right)^2 
           + \frac{1}{2\p\r}\left(J^{(1)} + J^{(2)}\right)\phi^{(0)} 
           \right. \\ \left.
       + \frac{1}{4\p\r}\left(J^{(1)}\eval{\Phi^{(B)}}_{M_{10}^{(1)}}
           + J^{(2)}\eval{\Phi^{(B)}}_{M_{10}^{(2)}}\right) \right\}\; .
\end{multline}
Since the solution~\eqref{massiveEOM} for $\Phi^{(B)}$ is linear in
the sources $J^{(i)}$, we see that by integrating out $\phi^{(B)}$ we
have generated a new term quadratic in the sources in the
ten-dimensional effective action. Furthermore, since the calculation
is purely classical and $\phi^{(B)}$ enters the action quadratically,
the process of integration is identical to substituting the solution
$\Phi^{(B)}(x)$ directly into the action~\eqref{scalarsubsS}. As we
have mentioned, the sources will, in general, be given in terms of other
fields of the theory, including perhaps $\phi^{(0)}$. The
corresponding equations of motion now arise by varying the fields in
the ten-dimensional effective action~\eqref{scalarten}. By integrating
out the massive modes we ensure that reducing the action is equivalent
to reducing the equations of motion. 

We would like to have an exact expression for the ten-dimensional
effective action in terms of the sources $J^{(i)}$ only. However, we
cannot give a closed form expression for the solution
$\Phi^{(B)}$. Nonetheless we can make an approximation. Since the
sources are assumed to vary slowly on the scale of the orbifold size,
we can write $\Phi^{(B)}$ as a momentum expansion in the inverse
wavelength of the sources. To zeroth order we can ignore the
ten-dimensional derivatives in the massive field equation of
motion~\eqref{massiveEOM}. Recalling that $\orbav{\Phi^{(B)}}=0$ and
that $\Phi^{(B)}$ is even under $Z_2$, we find the solution 
\begin{equation}
\label{approxsol}
\Phi^{(B)} = - \frac{\p\r}{12}\left[
      \left( 3\left(x^{11}/\p\r\right)^2
          - 6\left|x^{11}/\p\r\right|
          + 2 \right) J^{(1)}
      + \left( 3\left(x^{11}/\p\r\right)^2
          - 1 \right) J^{(2)}
      \right] + \ldots
\end{equation}
Here the dots represent higher-order terms in the momentum
expansion. They are of the form 
$f(x^{11}/\r)\r^{2n+1}\pt_{(10)}^{2n}J^{(i)}$
where $\pt_{(10)}^2$ is the ten-dimensional Laplacian. These
correspond to higher-dimension terms usually dropped in a Kaluza-Klein
reduction since they are suppressed by $(\r/\l)^{2n}$ where $\l$ is
the wavelength of the field in ten dimensions. They have been computed
in ref.~\cite{llo} for an explicit five-brane soliton solution of
M--theory on $S^1/Z_2$.

Substituting into the action~\eqref{scalarten},
keeping only the lowest dimension terms so that the
correction terms in the solution~\eqref{approxsol} can be dropped, we
find
\begin{equation}
\label{Sten}
   S_{10} = - 2\p\r \int_{M_{10}} \left\{
       \frac{1}{2}\left(\pt\phi^{(0)}\right)^2 
       + \frac{1}{2\p\r}\left(J^{(1)} + J^{(2)}\right)\phi^{(0)}
       - \frac{1}{24}\left( {J^{(1)}}^2 - J^{(1)}J^{(2)} 
           + {J^{(2)}}^2 \right) \right\}
\end{equation}
When performing the reduction of the full eleven-dimensional
description of the strongly coupled heterotic string we will always
find this characteristic form ${J^{(1)}}^2-J^{(1)}J^{(2)}+{J^{(2)}}^2$
appearing.  

In the discussion so far, we have glossed over one subtlety. We have
taken an example where the bulk field $\phi$ enters only quadratically
and couples linearly to the boundary sources. In general, the
situation will be more complicated. There may be non-linear sources
for $\phi$ both in the bulk and in the boundary. Fortunately, in the
strongly coupled string theory, we can in general treat these sources
perturbatively. The boundary sources are suppressed by a power of
the eleven dimensional gravitational coupling, namely $\k^{2/3}$, with
respect to the bulk, while the bulk sources are further suppressed. 
Thus there is a perturbative expansion of the reduced action in
$\r\k^{-2/9}$. Taking a scalar field example with an analogous
structure, we treat the massive mode as a small perturbation, and
expand the eleven-dimensional action as a power series in
$\phi^{(B)}$. To first non-trivial order in $\k$, the boundary sources
are independent of the massive field $\phi^{(B)}$. We then have the
approximate solution discussed above which, by analogy with the
strongly coupled string theory case, we will assume to be of order
$\k^{2/3}$. To this order, there is no contribution from bulk sources
for the massive mode. We then iterate, using this solution at linear
order in the boundary and bulk sources to calculate a corrected
solution $\Phi^{(B)}$, which will include pieces of higher order in
$\k$.  

Suppose we are interested only in an effective action to order
$\k^{4/3}$. To what order must we keep the solution? At first sight it
would appear we need to keep terms up to order $\k^{4/3}$. However, we have
the familiar result that substituting a solution to the equations of
motion into the action gives no contribution to linear order,
precisely because the solution is a point where the first-order
variation of the action vanishes. The leading behavior of the solution
$\Phi^{(B)}$ is of order $\k^{2/3}$. Thus, to obtain all the terms of
order $\k^{4/3}$, we need only substitute the leading order solution
and keep terms quadratic in $\Phi^{(B)}$. This corresponds to
substituting precisely the linearized solution~\eqref{massiveEOM}
and~\eqref{approxsol} we derived above.  

%%%%%%%%%%%%%%%%%%%%%%%%%%%%%%%%%%%%%%%%%%%%%%%%%%%%%%%%%%%%%%%%%%%%%%%%%%%%

\section{$F^4$, $F^2R^2$ and $R^4$ terms in the ten-dimensional effective
action}

We would now like to use the dimensional reduction procedure defined
above to reduce the eleven-dimensional description of the strongly
coupled $E_8\times E_8$ heterotic string to a ten-dimensional
theory. This can then be compared with the loop expansion of the
ten-dimensional effective action for the weakly coupled string. 

As we discussed in the introduction, we will not derive all possible
terms in the ten-dimensional theory. We will concentrate on terms up
to quartic order in the Riemann and gauge curvatures. Furthermore, we
will only explicitly consider the parity-odd quartic terms. This will
be equivalent to including only terms appearing up to order $\k^{4/3}$
in an expansion in the eleven-dimensional gravitational coupling. 
However, if required, the procedure could be used to calculate a much
larger class of terms.  

Witten and Ho\v{r}ava have argued that the low-energy effective action
for the strongly coupled heterotic string is described by
eleven-dimensional supergravity on a $S^1/Z_2$ orbifold with gauge
fields on each of the orbifold planes~\cite{hw1,hw2}. The resulting
action has an expansion in the eleven-dimensional coupling constant
$\k$, with terms appearing at increasing powers of $\k^{2/3}$. If we
write for the bosonic fields 
\begin{equation}
\label{action}
   S = S_0 + S_{\k^{2/3}} + S_{\k^{4/3}} + \ldots 
\end{equation}
then, in the upstairs picture, $S_0$ is the usual eleven-dimensional
supergravity theory, 
\begin{equation}
  \label{action0}
   S_0 = \frac{1}{2\k^2}\int_{M^{11}}\sqrt{-g}\left\{
           - R - \frac{1}{24}G_{IJKL}G^{IJKL}
           - \frac{\sqrt{2}}{1728}\e^{I_1...I_{11}}
               C_{I_1I_2I_3}G_{I_4...I_7}G_{I_8...I_{11}} \right\}\; ,
\end{equation}
where $G_{IJKL}=24\pt_{[I}C_{JKL]}$ is the field strength of the three-form
$C_{IJK}$. Under the $Z_2$ orbifold symmetry, $g_{AB}$, $g_{11\, 11}$ and
$C_{11AB}$ are even, while $g_{11A}$ and $C_{ABC}$ are odd. To next order,
the term $S_{\k^{2/3}}$ is localized purely on the fixed planes, and is given
by
\begin{multline}
   \label{action1}
   S_{\k^{2/3}} = - \frac{c}{8\pi\k^2}\left(\frac{\k}{4\pi}\right)^{2/3}
        \int_{M_{10}^{(1)}}\sqrt{-g}\;\left\{
           \tr(F^{(1)})^2 - \frac{1}{2}\tr R^2\right\} \\
        - \frac{c}{8\pi\k^2}\left(\frac{\k}{4\pi}\right)^{2/3}
           \int_{M_{10}^{(2)}}\sqrt{-g}\;\left\{
               \tr(F^{(2)})^2 - \frac{1}{2}\tr R^2\right\}\; .
\end{multline}
Here we have included the $R^2$ terms which were argued for
in~\cite{low1}. As discussed there, these terms cannot be fixed up to
the addition of terms quadratic in the Ricci tensor and scalar. Since
we will only be interested in Riemann-tensor terms in the reduced
ten-dimensional theory, here we keep only the Riemann-squared term,
the coefficient of which is fixed. 

In Ho\v{r}ava and Witten's original formulation of the theory, it was
argued that anomaly cancelation implied that the $c=1$ in the
action~\eqref{action}. However, subsequently Conrad has argued that
the correct factor should be $c=2^{-1/3}$~\cite{conrad} (see
also~\cite{deA}). Since we will mostly be interested in the form of
the final result rather than explicit coefficients, in what follows we
will keep $c$ general to allow for either possibilities.

We see that the presence of $S_{\k^{2/3}}$ introduces a source, localized on
the orbifold planes, to the gravitational equations of motion. One finds
\begin{multline}
\label{geom}
   R_{IJ} - \frac{1}{2}g_{IJ}R = - \frac{1}{24} \left(4G_{IKLM}{G_J}^{KLM} - 
       \frac{1}{2}g_{IJ}G_{KLMN}G^{KLMN}\right) \\
   -\frac{c}{2\pi}(\k /4\pi )^{2/3}\left(\d (x^{11})T_{IJ}^{(1)}
       +\d (x^{11}-\pi\r )T_{IJ}^{(2)}\right) \; ,
\end{multline}
where
\begin{equation}
\label{emt}
   T^{(i)}_{AB} = (g_{11,11})^{-1/2}\left\{
       \tr F_{AC}^{(i)}F_B^{(i)C} -
       \frac{1}{4}g_{AB}\tr(F^{(i)})^2
       - \frac{1}{2}\left(
           \tr R_{AC}{R_B}^C -\frac{1}{4}g_{AB}\tr )^2 \right) \right\}\; .
\end{equation}

In order to keep the action $S_0+S_{\k^{2/3}}$ supersymmetric, a
source must also be appended to the Bianchi identity for $G$. One finds
\begin{equation}
\label{Bianchi}
   \left(dG\right)_{11ABCD} = -\frac{c}{2\sqrt{2}\pi}
       \left(\frac{\k}{4\pi}\right)^{2/3} \left\{ 
          J^{(1)}_{ABCD}\d(x^{11}) + J^{(2)}_{ABCD} \d (x^{11}-\pi\r )
          \right\}
\end{equation}
with the sources $J^{(i)}$ defined as
\begin{equation}
\label{Jdef}
   J^{(i)}_{ABCD} = 6\left[ \tr F^{(i)}_{[AB}F^{(i)}_{CD]}
         - \frac{1}{2}\tr R_{[AB}R_{CD]} \right] 
      = \left[d \o^{(i)}_3\right]_{ABCD} \; .
\end{equation}
The three-forms $\o^{(i)}_3$ can be expressed in terms of the
Yang-Mills and Lorentz Chern-Simons forms $\o^{{\rm YM},(i)}_3$ and
$\o^{{\rm L}}_3$ as
\begin{equation}
  \o^{(i)}_3 = \o^{{\rm YM},(i)}_3 - \frac{1}{2}\o^{{\rm L}}_3 \; .
  \label{CS_def}
\end{equation}

As in the example in the previous section, the theory can also be
formulated in the downstairs picture. First, the zeroth-order action is
written with a factor $1/\k^2$ rather than $1/2\k^2$ since the
integration range is now restricted to $x^{11}=[0,\p\r]$. If we
assume, as is usual, that the variation of the bulk fields is taken to
be zero at the boundaries when calculating equations of motion, then
the $g$ and $C$ equation of motion have no contributions from the
boundaries. However, we must impose the effects of the modified
Bianchi identity. Recalling $G_{ABCD}$ is odd under the $Z_2$
symmetry, one can integrate the Bianchi identity over a small volume
intersecting the orbifold plane to get the equivalent boundary
conditions for $G$
\begin{equation}
\label{bdryG}
\begin{aligned}
   \left.G_{ABCD}\right|_{x^{11}=0} 
      &= -\frac{c}{4\sqrt{2}\pi}\left(\k/4\pi\right)^{2/3} J^{(1)}_{ABCD} \\
   \left.G_{ABCD}\right|_{x^{11}=\pi\r} 
      &= \frac{c}{4\sqrt{2}\pi}\left(\k/4\pi\right)^{2/3} J^{(2)}_{ABCD} \; .
\end{aligned}
\end{equation}
From the form of the equations of motion derived in the upstairs
picture, we know that the Einstein equation also has a source localized
on the boundaries. It is easy to show that this translates into a
boundary condition on the intrinsic curvature of the orbifold
planes. One finds
\begin{equation}
\label{bdryg}
\begin{aligned}
   K^{(1)}_{IJ} - \frac{1}{2}h^{(1)}_{IJ}K^{(1)} &= 
      -\frac{c}{2\pi}\left(\k/4\pi\right)^{2/3}T^{(1)}_{IJ} \\
   K^{(2)}_{IJ} - \frac{1}{2}h^{(2)}_{IJ}K^{(2)} &= 
      -\frac{c}{2\pi}\left(\k/4\pi\right)^{2/3}T^{(2)}_{IJ} \\
\end{aligned}
\end{equation}
where $T^{(i)}$ was defined in~\eqref{emt}, the intrinsic curvature
$K^{(i)}_{IJ}$ is given by~\cite{wald}
\begin{equation}
   K^{(i)}_{IJ} = h^{(i)K}_I\nabla_K n^{(i)}_{J}
\end{equation}
and $h^{(i)}_{IJ}=g_{IJ}-n^{(i)}_In^{(i)}_J$ is the induced metric on
the boundary and $n^{(i)}_I$ are the normal vectors. 

Finally, we should discuss the order $\k^{4/3}$ terms in the eleven-dimensional
theory which are relevant to our calculation. There are two types of such
terms, namely Green-Schwarz terms needed to cancel gravitational anomalies
in M--theory on $S^1/Z_2$ and $R^4$ terms which are paired
to the former by supersymmetry. Both types of terms are bulk terms and are
not specific to M--theory on $S^1/Z_2$ but rather are always present. 
The Green-Schwarz terms can be obtained from five-brane anomaly
cancelation~\cite{dlm,w5}, by comparison to type IIA string theory~\cite{vw}
or from gravitational anomaly cancelation in M--theory on
$S^1/Z_2$~\cite{hw2,deA,conrad}. All approaches lead to 
\begin{equation}
 S_{\k^{4/3},{\rm GS}} \propto \int_{M_{11}}C\wedge
     \left( -\frac{1}{8}\tr R\wedge R\wedge R\wedge R
     +\frac{1}{32}\tr R\wedge R\wedge\tr R\wedge R
     \right)\; .
\label{GS}
\end{equation}
The corresponding $R^4$ terms~\cite{R4} are given by
\begin{equation}
 S_{\k^{4/3},R^4} \propto \int_{M_{11}}\sqrt{-g}\,
                    t_8^{I_1\ldots I_8}\, t_8^{J_1\ldots J_8}R_{I_1I_2J_1J_2}
                    \ldots R_{I_7I_8J_7J_8}\; .
 \label{R4}
\end{equation}
A general definition of the tensor $t_8$ can be found
in~\cite{spr}. Acting on antisymmetric tensors $Y_{IJ}$, $Z_{IJ}$ it
takes the form 
\begin{multline}
 t_8^{I_1\ldots I_8}\, Y_{I_1I_2}Y_{I_3I_4}Z_{I_5I_6}Z_{I_7I_8} =
     -2\, Y_{IJ}Y^{IJ}Z_{KL}Z^{KL}-4\, Y_{IJ}Y_{KL}Z^{IJ}Z^{KL} \\
     +16\, Y_{IK}Y^{JK}Z^{IL}Z_{JL}+8\, Y_{IJ}Y_{KL}Z^{JK}Z^{LI}\; .
\end{multline}
Thus, for instance, we have in short hand notation
\begin{equation}
 t_8t_8R^4 = 6\, t_8(4\,\tr R^4 -\tr R^2\tr R^2 )\; .
\end{equation}
The latter result can be used to rewrite the $R^4$--terms~\eqref{R4} and
combine them with the Green-Schwarz terms~\eqref{GS} into the
expression, in the upstairs picture,
\begin{equation}
   S_{\k^{4/3}} = \frac{c'}{2\k^2}\left(\frac{\k}{4\pi}\right)^{4/3}
      \frac{1}{3\cdot 2^{11}\p^2} \int_{M_{11}} 
          \left(t^8-\frac{1}{\sqrt{2}}\e^{(11)}C\right)(4\,\tr R^4
          - \tr R^2 \tr R^2)
 \label{action2}
\end{equation}
where
\begin{equation}
   \e^{(11)}C = \e^{I_1\ldots I_8JKL}C_{JKL}\; .
\end{equation}
Again, since there is some debate over the coefficient in this
term~\cite{deA,conrad}, we have introduced a parameter $c'$. 
This form of the $\k^{4/3}$ terms is adapted to the supersymmetric invariant
combinations of the reduced ten-dimensional theory, as we will see below.
More specifically, the reduction of $t^8-\frac{1}{\sqrt{2}}\e^{(11)}C$
acting on a fourth power of the curvature leads to a $N=1$ supersymmetric
invariant in ten dimensions. This shows that a $D=11$ supersymmetric invariant
of fourth power in the curvature should at least contain either the terms
proportional to $\tr R^4$ or the terms proportional to $\tr R^2\tr R^2$ in
eq.~\eqref{action2}. Presumably, both terms are required by eleven-dimensional
supersymmetry so that eq.~\eqref{action2} precisely represents the
$D=11$ supersymmetric invariant.

We note that in general other explicit quartic terms are allowed. In
particular, there can be terms on the orbifold fixed planes. However,
these enter at a higher order in the $\k$ expansion. They have a
relative coefficient of $\k^{14/9}$, and so do not contribute to order
$\k^{4/3}$. Such terms are expected to contribute to the parity-even
quartic invariants, since they include a non-trivial dilaton
dependence. 

\bigskip

We are now ready to apply the reduction procedure, which we have explained in
section 2, in order to find the $F^4$, $F^2R^2$ and $R^4$ terms in a
ten-dimensional low-momentum limit of the above theory. We note at
this point that, since the sources for $G$ appear in the Bianchi
identity, they cannot be directly incorporated into the action as
stands. Nonetheless, the reduction procedure described in the previous
section can be simply generalized to include this case. The result is
that, as before, one simply substitutes the linear massive solution
for $G$ into the action to obtain the correct, dimensionally reduced,
effective action. To do the reduction, we split the bulk fields
according to 
\begin{equation}
\begin{aligned}
\label{splitup}
  C_{IJK} &= C_{IJK}^{(0)}+C_{IJK}^{(B)} \\
  G_{IJKL} &= G_{IJKL}^{(0)}+G_{IJKL}^{(B)} \\
  g_{IJ} &= g_{IJ}^{(0)}+g_{IJ}^{(B)} 
\end{aligned}
\end{equation}
into zero mode and massive background pieces. The former represent the actual
ten-dimensional degrees of freedom which must be $Z_2$--even
components of the eleven-dimensional fields. For these, we write
\begin{equation}
\begin{aligned}
\label{zeromodes}
  C^{(0)}_{AB11} &= \frac{1}{6}B_{AB} \\
  G_{ABC11}^{(0)} &= 3\partial_{[A}B_{BC]} \\
  {ds^{(0)}}^2 \equiv g^{(0)}_{IJ}dx^Idx^J &= e^{-2\f/3}g_{AB}dx^Adx^B
                        +e^{4\f/3}(dx^{11})^2\; ,
\end{aligned}
\end{equation}
where $B_{AB}$, $g_{AB}$ and $\f$ are $x^{11}$ independent and represent the
two-form field, the ten-dimensional metric and the dilaton, respectively.
The factor $e^{-2\f/3}$ in front of the ten-dimensional part of the metric
${ds^{(0)}}^2$ has been included for convenience to arrive at the
ten-dimensional string frame. 

As in section two, we also separate the sources for $G$ and $g$ into
massive and massless parts. The background fields $G_{IJKL}^{(B)}$ and
$g_{IJ}^{(B)}$ then include explicitly $x^{11}$-dependent pieces
needed to properly account for the source terms on the orbifold fixed
hyperplanes. Expanding to the order $\k^{2/3}$, they are determined in the
upstairs picture by the equations 
\begin{equation}
\label{upeom}
\begin{gathered}
  D_IG^{(B)IJKL} = 0 \\
  dG^{(B)}_{11ABCD} = -\frac{c}{2\sqrt{2}\pi}
      \left(\frac{\k}{4\pi}\right)^{2/3} 
      \left\{ J^{(1)}\d(x^{11}) + J^{(2)} \d (x^{11}-\pi\r )
      - \frac{1}{2\p\r}\left(J^{(1)}+J^{(2)}\right)\right\}_{ABCD}
      \\
\begin{split}
  R^{{\rm (lin)}}_{IJ} &=  
     \frac{1}{2}\left(D^2 g^{(B)}_{IJ} + D_ID_J g^{(B)} - D^KD_I g^{(B)}_{JK}
      - D^KD_Jg^{(B)}_{IK}\right) \\
     &= \frac{c}{2\pi}\left(\frac{\k}{4\pi}\right)^{2/3}\left\{
          \d (x^{11})S^{(1)}_{IJ}+
          \d (x^{11}-\pi\r ) S^{(2)}_{IJ}
          - \frac{1}{2\p\r} \left(S^{(1)}_{IJ}+S^{(2)}_{IJ}\right)
          \right\}
\end{split}
\end{gathered}
\end{equation}
where $S^{(i)}_{IJ}$ is given in terms of the energy momentum tensor
$T^{(i)}_{IJ}$, eq.~\eqref{emt}, as
\begin{equation}
 S^{(i)}_{IJ}=T^{(i)}_{IJ}-\frac{1}{9}g_{IJ}T^{(i)}\;,\qquad
 T^{(i)}=g^{IJ}T^{(i)}_{IJ}\; .
 \label{Sdef}
\end{equation}
To lowest order in the momentum expansion explained in section 2, we find
the solution~\cite{low1}
\begin{equation}
\label{upsol}
\begin{gathered}
\begin{aligned}
  G^{(B)}_{ABCD} &= -\frac{c}{4\sqrt{2}\pi}\left(\frac{\k}{4\pi}\right)^{2/3}
      \left\{ \e (x^{11})J^{(1)} - (x^{11}/\pi\r)(J^{(2)}+J^{(1)})
      \right\}_{ABCD} \\
  G^{(B)}_{ABC11} &= -\frac{c}{4\sqrt{2}\pi^2\r}\left(\frac{\k}{4\pi}
      \right)^{2/3}\left(\o^{(1)}_3+\o^{(2)}_3\right)_{ABC} 
\end{aligned} \\
  g^{(B)}_{IJ} = \frac{c\r}{12}\left(\frac{\k}{4\pi}\right)^{2/3}e^{2\f}\left\{
             \left( 3\left(x^{11}/\p\r\right)^2
                - 6\left|x^{11}/\p\r\right| 
                + 2 \right) S^{(1)}_{IJ}
             + \left( 3\left(x^{11}/\p\r\right)^2
                - 1 \right) S^{(2)}_{IJ} \right\}
\end{gathered}
\end{equation}
The currents $J^{(i)}$ are defined in eq.~\eqref{Jdef} and the explicit
form of $S^{(i)}_{IJ}$ can be read off from eq.~\eqref{Sdef} and \eqref{emt}.
Note that these expressions have the correct $Z_2$ symmetry properties; that
is, $G^{(B)}_{ABCD}$ is odd and $G^{(B)}_{ABC11}$, $g^{(B)}_{AB}$ and
$g^{(B)}_{11,11}$ are even while the off-diagonal entries $g^{(B)}_{A11}$
of the metric which are odd vanish since $S^{(i)}_{A11}=0$. The above
expressions satisfy the upstairs equations of motion~\eqref{upeom}. The
$\d$--function sources in these equations arise from the step-function
discontinuities of $G^{(B)}_{ABCD}$ and $\partial_{11}g^{(B)}_{IJ}$ at
$x^{11}=0,\p\r$. Eq.~\eqref{upsol} can, however, also be interpreted
as the solution in the boundary picture. In this case, $x^{11}$ is restricted
to $x^{11}\in [0,\p\r ]$ and the step function in the expression for
$G^{(B)}_{ABCD}$ and the modulus in the expression for $g_{IJ}^{(B)}$
become obsolete. Then the downstairs equations of motion, which can be obtained
from eq.~\eqref{upeom} by omitting the $\d$--function source terms, are
fulfilled and the boundary conditions~\eqref{bdryG} and \eqref{bdryg}
are properly matched.

\bigskip

Before we proceed to the computation of the higher-order terms, let us
first derive the effective ten-dimensional action to order $\k^{2/3}$
to settle our conventions. Inserting the fields specified by
eqs.~\eqref{splitup}, \eqref{zeromodes} and \eqref{upsol} into the action
\eqref{action}, \eqref{action0}, \eqref{action1} we obtain, to
order $\k^{2/3}$
\begin{equation}
 S_{10} = \frac{1}{2\k_{10}^2}\int_{M_{10}}\sqrt{-g}\, e^{-2\f}\left[-R+
          4(\partial\f )^2-\frac{1}{6}H^2-\frac{\a '}{4}\left(\tr (F^{(1)})^2
          +\tr (F^{(2)})^2\right)+\frac{\a '}{4}\tr R^2\right]\; ,
 \label{S01}
\end{equation}
where the three-form field strength $H_{ABC}$ is defined by
\begin{equation}
 H_{ABC} = 3\partial_{[A}B_{BC]}-\frac{\a '}{2\sqrt{2}}\left\{
     \o_3^{{\rm YM},(1)}+\o_3^{{\rm YM},(2)}-\o_3^{\rm L}
     \right\}_{ABC}\; .
 \label{Hdef}
\end{equation}
Here, we have made the identifications
\begin{equation}
 \k_{10}^2=\frac{\k^2}{2\p\r}\; ,\qquad
 \a ' = \frac{c}{2\p^2\r}\left(\frac{\k}{4\p}\right)^{2/3}\; .
 \label{ka}
\end{equation}
We recognize eq.~\eqref{S01} as the zero slope effective action of the
weakly coupled heterotic string to the first order in $\a '$. It is
known~\cite{susyR2} that the $R^2$ term in this action is required by
supersymmetry once the Lorentz Chern-Simons form $\o_3^{\rm L}$ is included
in the definition~\eqref{Hdef} of $H$. Such a term would not appear in the
dimensional reduction procedure unless it is explicitly included in the
boundary actions of the eleven-dimensional theory, as done in 
eq.~\eqref{action1}. This constitutes one rationale for the presence
of such terms in the eleven-dimensional theory, as pointed out in
ref.~\cite{low1}.

\bigskip

We are now going to calculate some order $\k^{4/3}$ corrections to the
action~\eqref{S01}, namely terms of the form $R^4$, $R^2F^2$, $F^4$ which
consist of four powers of curvatures and the corresponding Green-Schwarz
terms of the form $BR^4$, $BR^2F^2$, $BF^4$. From the eleven-dimensional
action and the field configuration we are going to use for the reduction,
we can already identify various sources for those terms. First, and
most obviously, such terms arise from the explicit eleven-dimensional $R^4$
terms given in eq.~\eqref{action2}. Those terms, however, cannot account
for the full spectrum of expected terms in ten dimensions and, in
particular, they do not lead to any such terms involving gauge fields.
At this point the $x^{11}$ dependent background fields $G^{(B)}_{ABCD}$ and
$g^{(B)}_{IJ}$ come into play. As can be seen from eq.~\eqref{upsol}, they
are of order $\k^{2/3}$ and are proportional to $\tr R^2$ and $\tr F^2$ so that
quadratic expressions of those backgrounds lead to terms of the right 
structure. Green-Schwarz terms in ten dimensions can only arise from
the eleven-dimensional ``Chern-Simons'' term $CGG$ where $C$ and $G$ are
replaced by $B$ and $G^{(B)}\sim\tr R^2\; ,\tr F^2$ respectively.
Pure curvature terms of the form $R^4$, $R^2F^2$, $F^4$, on the other hand,
result from three distinct sources, namely from the bulk curvature expanded
up to second order in the metric background $g^{(B)}_{IJ}$, from the
four-form kinetic term $\sqrt{-g}G^2$ with $G$ replaced by $G^{(B)}_{ABCD}$
and from the expansion of the boundary actions up to first order in the
metric background $g^{(B)}_{IJ}$ taken at $x^{11}=0,\p\r$. The explicit
calculation is performed by inserting the background~\eqref{upsol} into the
action specified by \eqref{action}, \eqref{action0}, \eqref{action1} and
\eqref{action2}. This leads to
\begin{equation}
   S_{10}(R^4,R^2F^2,F^4) = 
        \frac{c^2}{3\cdot 2^{16}\p^5\a'} \int_{M_{10}}\sqrt{g}\,
        \left( t_8-\frac{1}{2\sqrt{2}}\e^{(10)}B\right)W_8\; ,
 \label{S2}
\end{equation}
where
\begin{multline}
   W_8 = 8\left(\tr{F^{(1)}}^2 \tr{F^{(1)}}^2 
              - \tr{F^{(1)}}^2 \tr{F^{(2)}}^2
              + \tr{F^{(2)}}^2 \tr{F^{(2)}}^2 \right) \\
         - 4\,\tr R^2 \left(\tr{F^{(1)}}^2 + \tr{F^{(2)}}^2 \right) 
         + 2\, \tr R^2 \tr R^2 
         + \frac{c'}{c^2}\left(4\,\tr R^4 - \tr R^2 \tr R^2\right)
 \label{W8}
\end{multline}
and
\begin{equation}
 \e^{(10)}B=\e^{I_1\ldots I_8JK}B_{JK}\; .
\end{equation}
Let us discuss some basic properties of the above result. If $Y$ and $Z$
are curvatures the combination
\begin{equation}
 \left( t_8-\frac{1}{2\sqrt{2}}\e^{(10)}B\right) X\; ,\qquad
  X=\tr Y^2\tr Z^2\quad\mbox{or}\quad X=\tr Y^4
\end{equation}
constitutes an invariant under $N=1$ supersymmetry in ten
dimensions~\cite{rsw,ts}. Our result, eq.~\eqref{S2}, is expressed as a sum
of such supersymmetric invariants and, hence, is supersymmetric. The
appearance of these supersymmetric combinations, though expected, is by
no means a trivial consequence of the reduction process. While this is
true for the terms resulting from the explicit $R^4$ terms in the
eleven-dimensional theory which correspond to the last term in
parentheses in the polynomial $W_8$, eq.~\eqref{W8}, all other terms
result from various sources and include bulk and boundary
contributions as described in detail above. 

Comparing with the known form of the Green-Schwarz anomaly term, we
see that the $c$ and $c'$ coefficients must be related
\begin{equation}
   c' = c^2
\end{equation}
Furthermore, a careful consideration of the exact overall coefficient
of the Green-Schwarz term will also fix $c$, just as it would in the
full eleven-dimensional theory. We note that, comparing with the
expression for one-loop quartic curvature terms~\cite{ejl,aks1,aks2}
quoted by Abe {\em et al.}~\cite{aks2}, we have $c=1$. 
However, the main result here is that, by the somewhat complicated procedure
of integrating out the massive modes in the eleven-dimensional theory,
we have succeeded in reproducing the full supersymmetric invariants of
the ten-dimensional theory, up to parity-odd quartic terms.

%%%%%%%%%%%%%%%%%%%%%%%%%%%%%%%%%%%%%%%%%%%%%%%%%%%%%%%%%%%%%%%%%%%%%%%%%%%

\section{Conclusion}

We have seen that, when making a consistent dimensional reduction on a
$S^1/Z_2$ orbifold with sources on the orbifold fixed planes, it is
not possible to simply truncate and drop the massive Kaluza-Klein
modes. Fields on the fixed planes, even if they are assumed to be
independent of the circle coordinate, always provide a source for the
massive modes. Thus the massive modes do not decouple from the zero
mode fields, and integrating them out leads to new terms in the
effective action. When the sources on the fixed planes can be treated
perturbatively, this provides a new procedure for making a consistent
dimensional reduction.

One can use this procedure to calculate terms in
the ten-dimensional effective action of the strongly coupled
$E_8\times E_8$ heterotic string. In this paper, we have isolated the
terms which are quartic in the curvature and, in the string frame, are
independent of the dilaton. Such terms first appear at one loop in the
weakly coupled theory and include the Green-Schwarz terms. In the
dimensional reduction they come from two sources, both from explicit
quartic terms in eleven-dimensions and from integrating out the
massive modes. We find the full $N=1$ supersymmetric invariants in
ten-dimensions. Furthermore, these enter in the correct combination to
give the full anomaly-canceling Green-Schwarz terms. 

Unlike the case of the type II string where there is strong evidence
for a non-renormalization theorem for the corresponding parity-odd
quartic terms~\cite{R4,rt}, what we have done here is simply to show
how the weak-coupling one-loop terms arise from the reduction of the
strongly coupled theory, expanding up to $\k^{4/3}$. It is possible
that other terms in the eleven-dimensional theory lead to parity-odd
terms with non-trivial dilaton dependence. A good example is an
explicit quartic term in the orbifold fixed-plane action. However, we
know such terms cannot be supersymmetric and so we expect them not to
be present. In a sense it is more natural to reverse the argument and
use a non-renormalization constraint to exclude certain terms in the
eleven-dimensional theory.  (Something which is simpler than
investigating the supersymmetry of the eleven-dimensional theory
directly.) 

One set of terms which we do expect to be present in the
ten-dimensional action are the parity-even terms corresponding to
tree-level $\a'$ corrections, which have the characteristic
coefficient $\zeta(3)e^{-2\f}$. It is natural to ask how such terms
appear here. There are two obvious candidates. Explicit quartic terms
on the orbifold fixed planes can have the correct curvature structure,
but enter with the power $e^{-2\f/3}$, and so are non-perturbative
from a weakly coupled perspective, and cannot be the source. However,
as has been discussed in the context of type II theories
in~\cite{GGV,rt}, we can expect the bulk $R^4$ term in eleven
dimensions gets a correction when the theory is compactified on a finite
interval. This is essentially a one-loop Casimir effect. The term as
stands corresponds to a one-loop supergravity correction in an
infinite space. On a finite interval, the momentum modes become
quantized and this shifts the form of the one-loop $R^4$ term. The
authors of~\cite{GGV,rt} have shown how this leads to a term
corresponding to the tree-level $R^4$ term in type II theories and we
expect the same effect here.

Two further points are worth making. First, the dimensional reduction
procedure introduced here is a useful method for deriving higher-order
supersymmetric invariants. While we concentrated only on the quartic
curvature terms, the same calculation could also produce the
corresponding terms involving higher powers of $H$ and the dilaton, as
well as higher order fermion terms, needed to make the full
supersymmetric invariant. Calculation of such invariants directly in
ten dimensions is an extremely laborious task~\cite{rsw}. 

Secondly, the non-renormalization of the one-loop term provides part
of the explanation of why the strong and weakly coupled theories,
reduced to four dimensions on a Calabi-Yau manifold, have the same
form~\cite{low1}. As an expansion in $\k$, one finds that it is
precisely the one-loop terms which give the corrections to the
lowest order effective action in both limits. Thus, since the
ten-dimensional actions agree, the form of the correction in four
dimensions is the same in each case. The actual situation is a little
more complicated since in the strongly coupled theory one is never
really reducing a ten-dimensional action. The Calabi-Yau space is
actually smaller than the orbifold size. However, as discussed
in~\cite{low1}, for the leading corrections, the heavy modes of the
compactification do not contribute. Thus the form of the effective
action is, in fact, independent of the relative sizes of the
Calabi-Yau space and the orbifold. Consequently, all terms in the
four-dimensional effective action resulting from the one-loop
operators are of the same form in the strong and weakly coupled
limits. Nonetheless it is important to note that while, to this order,
the form of the actions is the same, the parameters are not, and this
can lead to quite different and interesting low-energy phenomenology.

%%%%%%%%%%%%%%%%%%%%%%%%%%%%%%%%%%%%%%%%%%%%%%%%%%%%%%%%%%%%%%%%%%%%%%%%%%%%

\vspace{0.4cm}

{\bf Acknowledgments} 
A.~L.~would like to thank Michael Faux for discussions. A.~L.~is
supported in part by a fellowship from Deutsche Forschungsgemeinschaft
(DFG). A.~L.~and B.~A.~O.~are supported in part by DOE under contract
No. DE-AC02-76-ER-03071. D.~W.~is supported in part by DOE under
contract No. DE-FG02-91ER40671. 

%%%%%%%%%%%%%%%%%%%%%%%%%%%%%%%%%%%%%%%%%%%%%%%%%%%%%%%%%%%%%%%%%%%%%%%%%%%%

%%%%%%%%%%%%%%%%%%%%%%%%%%%%%%%%%%%%%%%%%%%%%%%%%%%%%%%%%%%%%%%%%%%%%%%%%%%%

\end{document}